\documentclass[aps,prb,twocolumn,showpacs,preprintnumbers,amsmath,amssymb,groupedaddress]{revtex4-1}
\usepackage{mathptm} 
\usepackage{dcolumn}                    
\usepackage{bm}                        
\usepackage{graphicx}
\usepackage{times}
\begin{document}
\newcommand {\beq}{\begin{equation}}
\newcommand {\eeq}{\end{equation}}
\newcommand {\bea}{\begin{eqnarray}}
\newcommand {\eea}{\end{eqnarray}}
\newcommand {\nn}{\nonumber}
\newcommand {\bb}{\bibitem}
\newcommand{\et}{{\,\it et al\,\,}}
\newcommand{\CuB}{CuBiS$_{2}$}
\newcommand{\BiTe}{Bi$_{2}$Te$_{3}$ }

\title{Benefits of Carrier Pocket Anisotropy to Thermoelectric Performance: The case of $p$-type AgBiSe$_2$}

\author{David Parker, Andrew F. May and David J. Singh}
\address{Oak Ridge National Laboratory, 1 Bethel Valley Rd., Oak Ridge, TN 37831}

\date{\today}

\begin{abstract}
We study theoretically the effects of anisotropy on the thermoelectric performance of $p$-type AgBiSe$_2$.  We present an apparent realization of the thermoelectric benefits of one-dimensional ``plate-like" carrier pocket anisotropy in the valence band of this material.  Based on first principles calculations we find a substantial anisotropy in the electronic structure, likely favorable for thermoelectric performance, in the valence bands of the hexagonal phase of the silver chalcogenide thermoelectric AgBiSe$_2$, while the conduction bands are more isotropic, and in our experiments do not attain high performance.  AgBiSe$_2$ has already exhibited a $ZT$ value of 1.5 in a high-temperature disordered fcc phase, but room-temperature performance has not been demonstrated.   We develop a theory for the ability of anisotropy to decouple the density-of-states and conductivity effective masses, pointing out the influence of this effect in the high performance thermoelectrics Bi$_2$Te$_3$ and PbTe.  From our first principles and Boltzmann transport calculations we estimate the performance of $p$-type AgBiSe$_{2}$.
\end{abstract}
\maketitle

\section{Introduction}
  
Anisotropy is a substantial contributor to many phenomena of technological importance in condensed
matter physics.  For example, today's high performance magnets owe their performance in large part to a sizable magnetocrystalline anisotropy, deriving from spin-orbit coupling, the associated crystal field, and a 
non-cubic crystal symmetry.

Anisotropy is also important for thermoelectric performance, or the figure-of-merit $ZT$ (defined in Appendix A).  Very recently \cite{mecholsky}
Mecholsky {\it et al} found that band ``warping", or a non-analyticity of the effective mass tensor near the band edge, can have a substantial effect on the
electronic transport relevant for thermoelectric behavior.  In that work as well as the recent review by Shakouri \cite{shakouri} it was pointed out
that the existence of a band edge with different masses in different directions favors thermoelectric performance.  This was also
shown in Ref. \onlinecite{chen}, and the benefits of anisotropy to thermoelectric transport, with regards to FeAs$_{2}$ and LiRh$_2$O$_4$, were shown in Refs. \onlinecite{usui,arita}.   In fact this basic scenario, applied to artificial structures, had been foreseen by Hicks and Dresselhaus \cite{hicks1,hicks2} in works on thermoelectric transport in superlattices and nanowires.  It was also shown
\cite{parker_prl} that one can realize the beneficial effects of low-dimensional {\it electronic} structures even in isotropic materials.

Numerous authors have noted \cite{snyder,mahan,liu,wang,sootsman,lee,disalvo,wang2,ohta,snyder2,lalonde} the effect of 
carrier pocket degeneracy on thermoelectric performance, noting that such degeneracy improves electrical conductivity without sacrificing the Seebeck coefficient, both
indispensable components of a useful thermoelectric.  Such degeneracy is nearly universally rooted in the placement of a band extremum away from the $\Gamma$ point since the crystal symmetry then dictates the appearance of multiple carrier pockets.  This crystal symmetry also ensures that anisotropy in the electronic transport resulting from a single carrier pocket, which we term ``pocket anisotropy", does not sacrifice the {\it overall} electronic transport of the crystal.  Examples of such pocket anisotropy include the valence bands of the high performance thermoelectrics Bi$_2$Te$_3$ and PbTe.  The valence bands of both these materials contain band edges away from the $\Gamma$ point with three \cite{stordeur} (for Bi$_2$Te$_3$) and two \cite{cuff} (for PbTe) disparate effective masses in different directions.  However, the transport is isotropic in the basal plane for Bi$_2$Te$_3$  and for all three dimensions in cubic PbTe.  Our calculations described below find essentially isotropic {\it overall} electrical transport for $p$-type AgBiSe$_2$ despite substantial {\it carrier pocket} anisotropy.

In addition, the benefits of mass anisotropy increase as the dimensionality of the electronic structure decreases.  For example, an isotropic three dimensional parabolic band has a $\sqrt{E}$ dependence of the density-of-states $N(E)$ on energy $E$, a two-dimensional parabolic band has a step-function behavior, and a fully one-dimensional band has an $E^{-1/2}$ behavior, implying a diverging density-of-states (DOS) at the band edge.  Note that a one-dimensional band takes on a ``sheet" or ``plate-like" isoenergy or Fermi surface.  While in real materials such a DOS divergence does not occur due to a lack of complete one-dimensionality, it is clear there are particular benefits of such a one-dimensional structure.  

In this work we present calculations of such one-dimensional ``plate-like" carrier pocket anisotropy in the valence band of the silver chalcogenide semiconductor
AgBiSe$_2$.  This material crystallizes in a hexagonal structure at room temperature, with successive transitions to a rhombohedral structure at 420 K and to a disordered
fcc structure at about 600 K.  Recent work \cite{pan,xiao1,xiao2} shows a $ZT$ value for $n$-type AgBiSe$_2$ of 1.5 just above the cubic-rhombohedral phase transition.  We find from theory that good room-temperature $p$-type thermoelectric performance may be possible, estimating a $p$-type $ZT$ value of 0.4-0.7, considering only electronic optimization.

Ordinarily one-dimensional electronic structures are associated with electronic instabilities such as charge or spin-density waves.  These instabilities typically lead to insulating ground state behavior.  Here, by contrast, we have a one-dimensional electronic structure \cite{usui} associated with a {\it metallic} ground state, that we will show to be highly beneficial for thermoelectric performance.  

We observe that very few materials have ZT $\sim$ 1 in the important temperature range between 200 and 350 K, where many heating and cooling applications exist.  Substantial efforts to find a substitute for the prototypical thermoelectric Bi$_2$Te$_3$, which shows a $ZT$ of unity around room temperature, have been made for nearly 50 years.  Hence the finding of a candidate for such performance levels, assuming full optimization, in this temperature range is of substantial importance.   This remains true even despite the difficulty, described below, of doping into the valence band of AgBiSe$_2$.  We note that both Bi$_2$Te$_3$ and AgBiSe$_2$ contain expensive elements (Tellurium and Silver).  However, Silver is more available, with worldwide annual production is 26,000 tons, compared to Tellurium, with annual production of 100 tons \cite{usgs}.  This parallels the 75 times larger mass abundance \cite{crc} in the Earth's crust of Silver relative to Tellurium.

The remainder of this paper is organized as follows: in Section II, we present our experimental procedure; in Section III we present our experimental data, along with a comparison of the calculated $n$-type and $p$-type electronic structures; in Section IV, we present a more general consideration of anisotropy, followed in Section V with our first principles theoretical results and 
our estimate of $ZT$, and our conclusions in Section VI.  Details on the estimation of $ZT$ are presented in Appendix A, and a demonstration of the isotropy of the Seebeck coefficient
within an effective mass approach is given in Appendix B.

\section{Experimental Procedure and Efforts to Attain $p$-Type Doping}

AgBiSe$_2$ samples were prepared by reacting high-purity elements in evacuated silica ampoules.  For the `as-grown' sample, solidification occurred slowly (1 deg/hr) to promote large grains.  The sample was then cooled at 5deg/hr between 310$^{\circ}$C and 260$^{\circ}$C, followed by a 48h anneal at 260$^{\circ}$C to minimize defects and reduce residual stresses. Due to the formation of voids during growth and and cracks during processing, only relatively small pieces were obtained.  These difficulties lead to the synthesis of polycrystalline samples, which also allow for a more rapid screening of potential dopants.  Following reaction in the melt, samples were ground in a He glove box and hot-pressed in a graphite die at temperatures near 450$^{\circ}$C with a pressure of approximately 10,000 psi.  This process resulted in samples with geometric densities of approximately 95\% of the theoretical density.  

The Seebeck coefficient was used as a screening tool to examine the influence of the various substitutions and processing conditions.  The Seebeck and Hall coefficients were measured in a Quantum Design Physical Property Measurement System, using the Thermal Transport and Resistivity Options, respectively.  Thermal measurements were performed using gold-coated copper leads attached to the sample with H20E Epo-Tek silver epoxy.  Hall data were obtained using a standard four-wire configuration, with 0.0508 mm Pt wires spot welded to the sample and maximum fields of $\pm$6T were employed; electrical resistivity was collected during the same measurement as the Seebeck coefficient and thermal conductivity.  Scanning electron microscopy (SEM) and energy dispersive spectroscopy (EDS) measurements were performed in a Hitachi TM-3000 microscope equipped with a Bruker Quantax 70 EDS system.

In order to manipulate the electrical properties, polycrystalline samples containing S, Te, and Pb were made.  Similarly, a sample of nominal composition Ag$_{1.1}$Bi$_{0.9}$Se$_2$ was produced from the melt to examine the potential for instrinsic hole doping via bismuth deficiency.  SEM coupled with EDS revealed a Ag-rich phase at apparent grain boundaries in this Ag$_{1.1}$Bi$_{0.9}$Se$_2$ sample, which indicates the tendency of the phase to form near the stoichiometric composition AgBiSe$_2$.  This sample was not considered further due to our desire to probe, as much as possible, only the intrinsic properties of AgBiSe$_2$.

In our experimental results, AgBiSe$_2$ naturally forms $n$-type.  As-grown AgBiSe$_2$ was found to have a Hall carrier density $n_H$ near 2.3$\times$10$^{19}$cm$^{-3}$ at room temperature, and little temperature variation was observed down to 25K.  At 300\,K, this as-grown sample had a Seebeck coefficient of -138$\mu$V/K  and a Hall mobility of $\mu_H$=45cm$^2$/V/s.  Our polycrystalline AgBiSe$_2$ had a Seebeck coefficient of $\approx$-400$\mu$V/K at 300\,K, and an electrical resistivity that increased with decreasing temperature.  These results suggest that our polycrystalline AgBiSe$_2$ is very lightly doped, or nearly an intrinsic semiconductor.  Thus, AgBiSe$_2$ can be formed near the insulating limit, which makes obtaining $p$-type conduction more plausible.

Pb has previously been shown\cite{pan} to induce $p$-type conduction in AgBiSe$_2$.  However, our Pb-doped sample of nominal composition AgBi$_{0.95}$Pb$_{0.05}$Se$_2$ had a Seebeck coefficient of $\approx$-95$\mu$V/K, suggesting a higher electron concentration than the as-grown sample or the polycrystalline AgBiSe$_2$.  The apparent increase in free electrons with a $p$-type dopant is surprising, and may be related to inhomogeneities within the sample.  However, the absence of $p$-type conduction at low Pb concentrations was reported by Pan et al \cite{pan}, though they observed a cross-over to $p$-type conduction near $x=0.02$.  We did not continue to investigate Pb-doping after this initial finding.  Instead, we briefly examined the influence of S and Te substitutions.  

The $n$-type behavior of AgBiSe$_2$ may be caused by Se vacancies.  To gain insight into this potential mechanism, we considered S and Te substitutions for Se.  Interestingly, our sample of nominal composition AgBiSe$_{1.8}$Te$_{0.2}$ had a room temperature Seebeck coefficient of -475$\mu$V/K and resitivity that increased with decreasing $T$.  Compared to the -400$\mu$V/K observed for the undoped, polycrystalline sample, this would suggest that Te substitution drives the systems towards a charge balanced state due to its lower vapor pressure (fewer anion vacancies).  However, we remain cautious because AgBiTe$_2$ is also naturally $n$-type\cite{Sakakibara}.  Also, Seebeck coefficient measurements become more difficult and absolute errors increase for resistive samples.  Yet, sulfur substitution pushed the system in the opposite direction (more metallic), consistent with a relative increase in elemental vapor pressure.  Unfortunately, there was no trend in these data and the more heavily substituted sample had a larger 300 K $n$-type Seebeck coefficient (-87$\mu$V/K for 5\% S, and -175$\mu$V/K for 10\%S).  As such, we cannot draw any conclusions regarding the sulfur substitution, though we remain optimistic about coupling Te substitution with non-isoelectronic dopants.  

A detailed study of the defect chemistry in these materials could provide the insights required to achieve the desired levels of $p$-type conduction.  A more complete picture would require additional experimental and theoretical efforts, considering the variety of compositions in this family of I-V-VI$_2$ chalcogenides. 

The large Seebeck coefficients and semiconducting behavior of the resistivity of our polycrystalline samples thus suggest that an adequate $p$-type dopant might be found.  We also note that in Refs. \onlinecite{xiao1,xiao2} $p$-type behavior - i.e. positive Seebeck coefficient  - was observed in the low-temperature hexagonal phase, so our results should not be taken to imply that $p$-type doping is impossible to achieve in AgBiSe$_2$, merely that more involved effort will likely be required to achieve this.   In the literature there are numerous semiconductors such as CrSi$_{2}$ and Bi$_{2}$Se$_{3}$ \cite{white,hor} which were originally found to exhibit a strong doping type preference, which were later found possible to dope both $n$-type and $p$-type under the proper circumstances.  Since AgBiSe$_2$ has not undergone much experimental study to date, we think it likely that the difficulty in attaining $p$-type doping will be successfully addressed by future efforts.  We note finally that the relatively narrow calculated band gap of 0.52 eV (see Section IV) argues in favor of the likelihood of attaining both doping types. 

\section{Experimental Data and Theoretical Comparison of $n$-type and $p$-type} 

In Figure 1 we show the T-dependent Seebeck coefficient, resistivity and thermal conductivity for the as-grown sample, along with (left panel) our first principles calculation of the Seebeck coefficient for this sample.   The measured Hall number, -0.27 cm$^{3}$/Coul, would correspond to an electron concentration $n$ of 2.3 $\times$ 10$^{19}$ cm$^{-3}$ in the case of an isotropic parabolic band, although Fig. 2  (top) depicts a conduction band electronic structure significantly differing from isotropic, so that the chemical and Hall-inferred carrier concentrations may differ. We find a good fit at a {\it chemical} carrier concentration $n$ of 2.52 $\times$ 10$^{19}$ cm$^{-3}$ indicating the accuracy of our theoretical approach, which uses the constant scattering time approximation \cite{parker_PbSe}.
\begin{figure} [h!]
\includegraphics[width=7cm,angle=90]{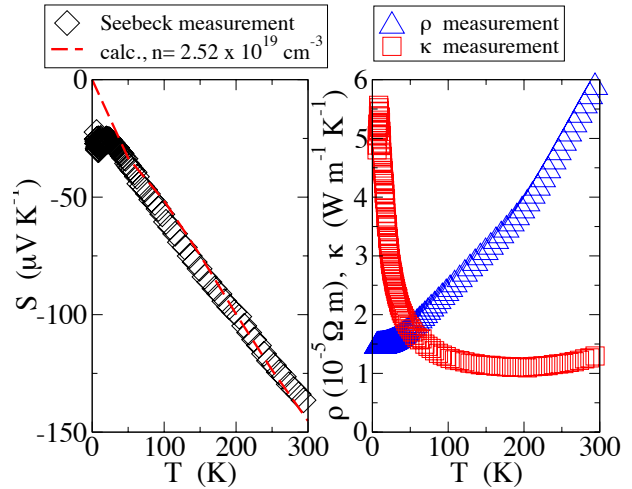}
\caption{  (Left panel) The measured and calculated thermopower of AgBiSe$_{2}$; right panel, the measured resistivity (blue triangles) and thermal conductivity (red squares).}
\end{figure}

The right hand panel of Fig. 1 shows the thermal conductivity and resistivity measurements.  We see that the thermal conductivity $\kappa$ of AgBiSe$_{2}$ is extremely low - it reaches a minimum value of 1.1 W/m-K around 200 K, indicating low lattice thermal conductivity - a key parameter of a useful thermoelectric. In fact, using the resistivity data and the Wiedemann-Franz relationship we find the lattice term at 200 K to be just 1.0 W/m-K, which is lower than that of the better-known high performance thermoelectrics such as Bi$_{2}$Te$_{3}$ and PbTe.   This value is slightly higher than that found by Nielsen {\it et al} for rhombohedral AgBiSe$_2$ \cite{nielsen}, presumably due to the difference in physical structure.  The slight upturn of $\kappa$ above 200 K likely reflects radiative effects not corrected for here.

The resistivity data shows the Fermi-liquid T$^{2}$ dependence at the very lowest temperatures below 50 K, crossing over to T-linear behavior for temperatures above 100 K.  This T-linear behavior is characteristic of electron-phonon scattering and allows us to estimate the electron phonon coupling constant $\lambda$ for this sample. We will later use this to estimate the performance of $p$-type AgBiSe$_{2}$.  From our theoretical calculations we find the plasma frequency squared $\omega_{p}^{2}$ at the modeled doping level to be 0.096 eV$^{2}$. When we combine this with our resistivity data and the theoretical relationship connecting $\lambda$, $\omega_P$, and the resistivity from Ref. \onlinecite{allen}, we find an electron-phonon coupling constant of 0.49, and an associated electronic scattering time of 8 $\times 10^{-15}$ s.

Although we were able to able to obtain significant $n$-type doping levels, our data suggest that in this temperature range $n$-type AgBiSe$_2$ is not likely to be a high performance thermoelectric.  The 300 K power factor $S^{2}\sigma$ for our as-grown sample is just 0.3 mW/m-K$^{2}$, or less than 10 percent of the value of optimized Bi$_2$Te$_2$.  Although one might achieve some gain in the power factor in a more lightly doped sample, this will not likely raise 300 K $ZT$ substantially from the $\sim 0.1$ value achieved here (a rough estimate finds optimized $ZT$ values of less than 0.2).

It is of interest to understand the reason for this.  Presented in Figure 2 (top) is a plot of the first-principles calculated isoenergy surface for $n$-type AgBiSe$_2$ for a doping $n=1.23 \times 10^{20}$ cm$^{-3}$.  The plot depicts a cylindrical body whose width is quite comparable to its height.  While the cylindrical surface is suggestive of two-dimensionality and hence mass anisotropy, the {\it shape} does not lend itself to a large surface to volume ratio, which we have argued elsewhere \cite{chen} to be favorable for high $ZT$.  The structure is also substantially lacking in degeneracy (technically there is a two-fold degeneracy as the band edge is at the A point).  Given these factors and the lack of the favorable complexity described in Ref. \onlinecite{shi}, it is perhaps not surprising that the performance levels of hexagonal $n$-type AgBiSe$_2$ are comparatively low.
\begin{figure} [h!]
\includegraphics[width=7.8cm]{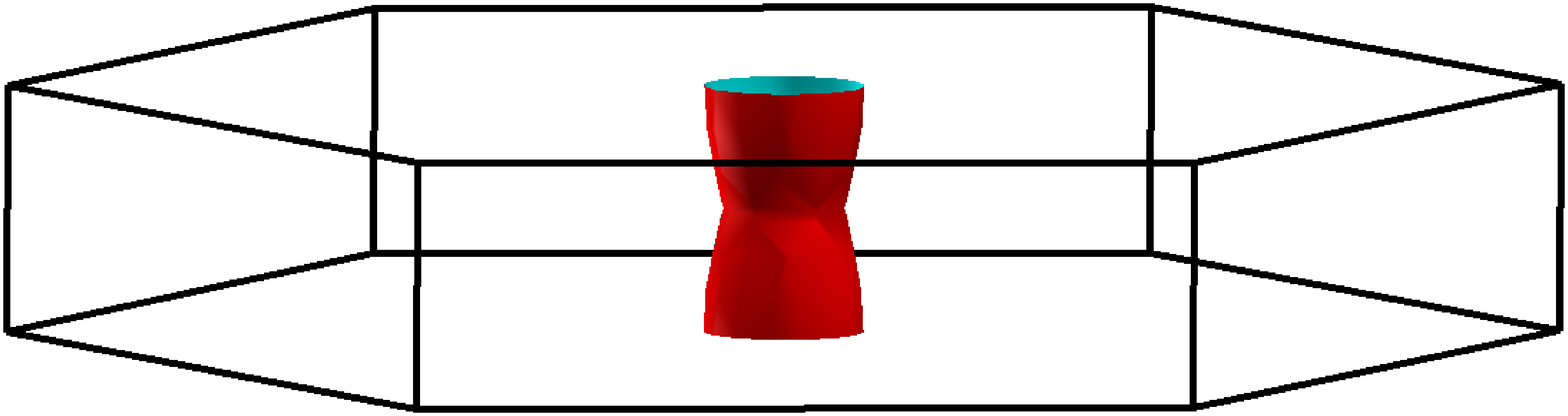}
\includegraphics[width=7cm]{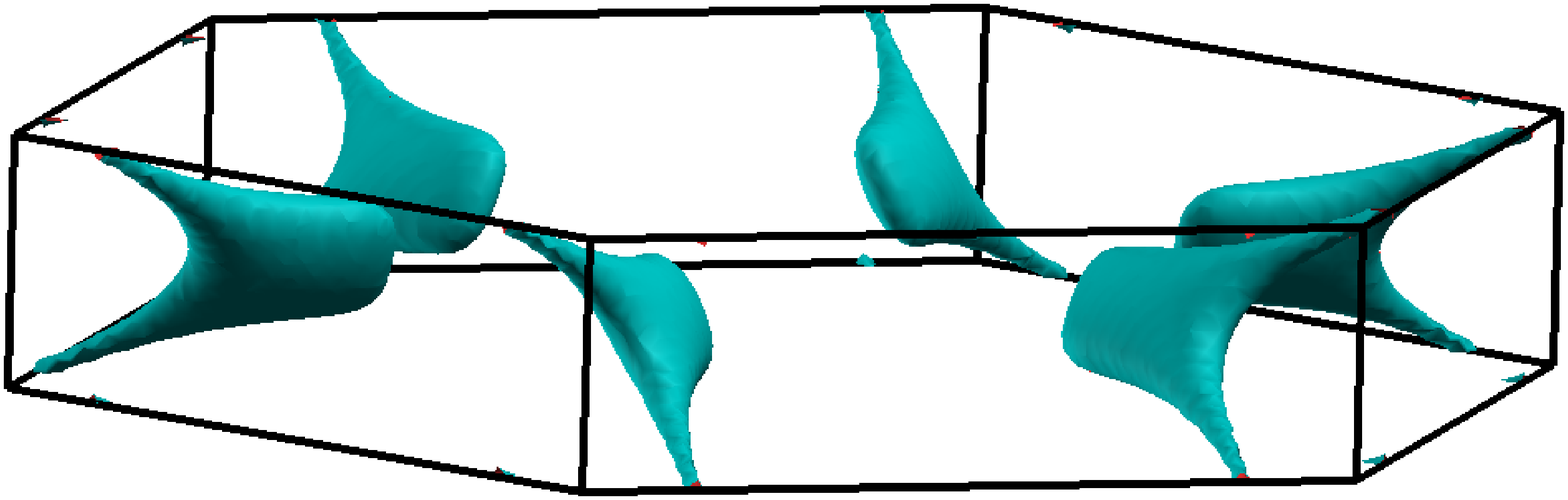}
\caption{  The calculated Fermi surface of hexagonal AgBiSe$_{2}$ at an electron doping $ n= 1.23 \times $ 10$^{20}$cm$^{-3}$) (top) and the same carrier concentration for $p$-type (bottom).  The bottom plot is rotated slightly to better depict the degeneracy and mass anisotropy.}
\end{figure}

The situation is rather different for $p$-type, however.
Figure 2 (bottom) presents a plot of the isoenergy surface for $p$-type AgBiSe$_2$ at the same carrier concentration as depicted for $n$-type.  This doping is likely near the optimal $p$-type doping for room-temperature $ZT$. The Fermi surface consists of six platelike structures that exhibit the feature mentioned in the Introduction: a perpendicular mass much less than the parallel mass.  In fact, in the plane of the ``plate", the effective mass is approximately 5 times the perpendicular mass.  There are additionally highly elongated features which stretch out to a secondary band maximum at a non-symmetry point on the zone surface.   We also note the six-fold degeneracy, a combination of the hexagonal symmetry and the position of the VBM away from $\Gamma$, and away from high-symmetry zone boundary positions.   Finally, one observes the {\it orientation} of the ``plates'': they are located at an angle to the c-axis, not parallel to it, so that the favorable transport properties of this feature will extend to the $c$-axis transport as well.  We will see that $p$-type transport in this material is expected to be virtually isotropic.

In the next section we give a general theoretical description of the benefits of carrier pocket anisotropy in enhancing thermoelectric performance, as appear to be active in $p$-type AgBiSe$_2$, specifically the role of such pocket anisotropy in decoupling the {\it density-of-states} or thermopower effective mass from the {\it conductivity} effective mass.

\section{Theory of role of pocket anisotropy in decoupling density-of-states and conductivity effective masses}

Here we show that any semiconductor band edge pocket anisotropy that can be expressed in terms of an ellipsoidal effective mass tensor 
has substantial benefits to thermoelectric performance, provided that the band edge is not located at the $\Gamma$ point and the crystal obeys a 
certain minimum symmetry.  Larger anisotropies, in addition to the effects of band degeneracy \cite{snyder}, are shown to be more beneficial in this regard.  

We begin with the canonical expressions for the temperature and chemical-potential-dependent thermopower tensor $S_{\alpha\alpha}(T,\mu)$ and the electrical conductivity tensor $\sigma_{\alpha\alpha}(T,\mu)$, which we reproduce here from Ref. \onlinecite{zhang_singh}. For simplicity we assume that both are diagonal in the spatial indices $\alpha$ and also assume only one band contributes to transport.  Then we have:
\bea
S_{\alpha\alpha}(T,\mu) &=& \frac{\nu_{\alpha\alpha}(T,\mu)}{\sigma_{\alpha\alpha}(T,\mu)}
\eea
with 
\bea
\sigma_{\alpha\alpha}(T,\mu)&=& -\int \sigma_{\alpha\alpha}(E)\frac{\partial f (T,E)}{\partial E} dE
\eea
and
\bea
\nu_{\alpha,\alpha}(T,\mu) &=& \frac{-1}{eT}\int \sigma_{\alpha\alpha}(E) (E-\mu) \frac{\partial f(T,E)}{\partial E} dE
\eea
Here $\sigma_{\alpha,\beta}(E) $ is the transport function which is written as
\bea
\sigma_{\alpha\alpha}(E) &=& e^2\sum_{i,{\bf k}}v^2_{{\bf k},\alpha,i}\tau_{i,{\bf k}}\delta(E-E_{i,{\bf k}})
\eea
and $f$ is the Fermi function.  Note that for an anisotropic parabolic band, in which the relaxation time $\tau$ depends on ${\bf k}$ through $E_{i,{\bf k}}$, the above expression (Eq. 4) for the transport function
can be written as
\bea
\sigma_{\alpha,\alpha}(E) &=& e^2 v^2_{\alpha}(E)\tau_{\alpha}(E) N(E)
\eea
Here the effective mass anisotropy is incorporated into the directional index $\alpha$.

As described in Appendix B, this effective mass anisotropy nonetheless yields an isotropic Seebeck coefficient, so long as the bands are taken
as parabolic, only carriers of one sign contribute to transport, and the scattering time is taken as depending only on energy, not on direction.
We note also that, though difficult to obtain in practice, a highly anisotropic Seebeck coefficient is of interest in the study of transverse thermoelectric effects \cite{goldsmid}.

We now analyze this problem through the concept of effective mass.  From standard references \cite{ashcroft} the electrical conductivity tensor may be written as
\bea
{\bf \sigma} &=& \tau \sum \frac{d^3 k}{4\pi^3}{\bf M}^{-1}
\eea
where ${\bf M}^{-1}$ is an effective mass tensor and the sum is taken over occupied levels.
The standard electronic conductivity
formula (where we have retained the directional indices) can then be written as
\bea
\sigma_{\alpha,\alpha}& = & \frac{ne^2 \tau_\alpha}{m_{\sigma,\alpha}}
\eea
and contains a conductivity effective mass $m_{\sigma,\alpha}$ to which the conductivity is inversely proportional.  For high electrical conductivity (a prerequisite for good thermoelectric performance) one therefore
desires small effective masses, irrespective of the scattering time $\tau_i$.  

The situation is very different for the thermopower $S$. In the fully degenerate limit the Mott formula gives the thermopower
as 
\bea
S_{\alpha,\alpha}(n,T) &=& \frac{\pi^2}{3}\frac{k_B}{e}k_B T \left(\frac{d\log(\sigma_{\alpha\alpha}(E))}{dE}\right)|_{E=E_{F}}
\eea

For an anisotropic parabolic band, one has (from Appendix B) that
\bea
\sigma_{\alpha\alpha}(E) &=&  g_{3,\alpha}(m_x,m_y,m_z)E^{3/2}
\eea
where E is measured from the band edge and $g_3$ is a function of the band masses.  The logarithmic derivative in the above equation is simply 3/2E$_{F}$.  $E_{F}$ may be rewritten
in terms of the carrier concentration $n$ and effective masses $m_x,m_y$ and $m_z$ by noting that $n$ can be written as (using the formula for the volume of an ellipsoid and including a factor of 2 for spin degeneracy)
\bea
n=\frac{2}{(2\pi)^3}\frac{4\pi}{3}k_{F,x}k_{F,y}k_{F,z}
\eea
where $\hbar k_{F,i}=\sqrt{2m_{i}E_{F}}$ with $m_{i}$ the effective mass in direction $i$, yielding
\bea
n=\frac{1}{3\pi^2\hbar^3}\sqrt{8m_x m_y m_z}E_{F}^{3/2}
\eea
Combining the above equations and rewriting $(m_x m_y m_z)^{1/3}$ as $m_{DOS}$, we finally have
\bea
S(T,n) &=&\frac{4\pi ^2k_B^2 }{eh^2}m_{DOS} T\left( {\frac{4\pi }{3n}} 
\right)^{2 / 3}
\eea
For the thermopower larger effective DOS masses are beneficial, which runs completely counter
to the benefit to the conductivity of smaller masses.  This is one major fundamental problem that must be overcome
to achieve high thermoelectric performance.

However, as suggested by the notation, {\it there is no particular reason that $m_{\sigma}$ and $m_{DOS}$ must be equal.}  They are in fact equal only
for a single isotropic parabolic band.
This distinction is well-known for the case of band degeneracy, where a band edge of degeneracy $N$ enhances \cite{snyder} $m_{DOS}$  by a factor $N^{2/3}$ without affecting $m_{\sigma}$.  Here we explore in addition the effect of anisotropy on the
relation of these two masses.  We find that anisotropy allows for small {\it conductivity} effective masses $m_{\sigma}$, heightening the electronic conductivity,
but large {\it density-of-states} effective masses $m_{DOS}$, enhancing the thermopower.

One can see the reason for this quite easily.  For simplicity, (the effect is similar for three distinct masses) consider an anisotropic parabolic band edge represented by a three-dimensional ellipsoid of revolution, characterized by two effective
masses, a radial mass $m_{\parallel}$ and a longitudinal mass $m_{\perp}$.  The DOS effective mass for this situation is given by $m_{DOS}$ = ($m_{\parallel}^2 m_{\perp})^{1/3}$.  The conductivity effective masses
in each of the two directions are given by $m_{\parallel}$ and $m_{\perp}$.  However, for a crystal in which the band edge is not at the $\Gamma$, or similarly low-degeneracy, point, there will generally be some band edge degeneracy which respects the crystal symmetry.  As depicted in Ref. \onlinecite{parker_prl}, this means that even an anisotropic band edge can result in isotropic transport in a cubic material.  The same general idea applies for planar transport for a layered material such as Bi$_{2}$Te$_3$.   One then sees that the (isotropic) conductivity effective mass is given by 
\bea
3/m_{\sigma} &=& 2/m_{\parallel} + 1/m_{\perp}
\eea
so that 
\bea
m_{\sigma} &=& \frac{3m_{\parallel} m_{\perp}}{2m_{\perp}+m_{\parallel}}
\eea
As is clear from the expression, this conductivity mass $m_{\sigma}$ has a rather different dependence on $m_{\perp}$ and $m_{\parallel}$ than the density-of-states mass $m^{}_{DOS}$.  Indeed, the ratio $m_{DOS}/m_{\sigma}$, which is effectively a figure-of-merit for the effects of electronic anisotropy, is simply
\bea
m^{}_{DOS}/m^{}_{\sigma} &=& \frac{2m_{\perp}+m_{\parallel}}{3(m_{\parallel}m^2_{\perp})^{1/3}} \\
&=& (2/3)r^{1/3}+(1/3)r^{-2/3}
\eea
In the last expression we have expressed $m^{}_{DOS}/m^{}_{\sigma}$ in terms of $r$ where $r$ is the mass ratio $m_{\perp}/m_{\parallel}$.  Now, this $m_{DOS}/m_{\sigma}$ ratio can be substantially different from unity if $m_{\parallel}$ and $m_{\perp}$ are very different.  For a case where $m_{\perp} = 24 m_{\parallel}$ (a value present in the conduction band of GeTe \cite{chen}, this
\begin{figure}
\vspace{-1.7cm}
\includegraphics[width=8cm]{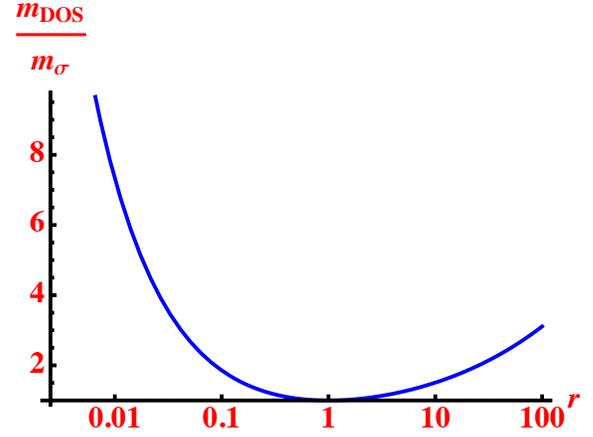}
\vspace{-2cm}
\caption{  The ratio $m_{DOS}/m_{\sigma}$.  Here $r=m_{\perp}/m_{\parallel}$.}
\end{figure}
ratio approaches the value 2.  Additionally, in the opposite limit, in which the band edge forms a flattened, highly prolate, rather than oblate, spheroid, still larger effects are present.  Figure 3 depicts a plot of the DOS to conductivity mass ratio as a function of the ratio $r=m_{\perp}/m_{\parallel}$.  For this ratio = 1/100  one finds a DOS to conductivity mass ratio of nearly 10, which is clearly beneficial in enhancing both thermopower and electrical conductivity.

As examples of these concepts, we consider the $p$-type band masses in the high performance thermoelectrics Bi$_2$Te$_3$ and PbTe.  The band edge in $p$-type Bi$_{2}$Te$_{3}$ has three distinct masses \cite{stordeur}, which take the values of 0.064 $m_{0}$, 0.196 $m_{0}$, and 0.73 $m_{0}$, where $m_{0}$ is the free electron mass.  These yield a band edge $m_{DOS}$ of 0.209 $m_{0}$ and $m_{\sigma}$ of 0.136 \cite{footnote} so that their ratio is over 50 percent enhanced relative to an isotropic band edge.  When one includes the six-fold degeneracy of the band edge in the DOS mass the effective mass ratio becomes 3.87, which is surely a major contributor to the performance of this material.  Similarly, the L-point band edge in PbTe contains a radial mass \cite{cuff} of 0.022 $m_{0}$ and longitudinal mass 0.31 $m_{0}$, yielding an $m_{DOS}/m_{\sigma}$ of 1.66, or 4.19 if the fourfold L-point degeneracy is included, which likely contributes to the exceptional thermoelectric performance \cite{kanat} of this material.

\section{Possible realization of beneficial effects of carrier pocket anisotropy: $p$-type {AgBiSe$_{2}$}}  

We focus here on the low temperature hexagonal phase of AgBiSe$_2$, calculating its properties with the first principles density functional theory code WIEN2K \cite{wien}, as well as Boltzmann transport properties via the Boltztrap code \cite{boltz}, within the generalized gradient approximation (GGA) of Perdew, Burke and Ernzerhof \cite{perdew}.  For these calculations we use a modification  of the GGA known as a modified Becke Johnson potential \cite{tran} which gives accurate band gaps \cite{koller,singh_tb} , a matter of great importance for the transport properties.  The linearized augmented plane wave (LAPW) basis was used, with LAPW sphere radii of 2.5 Bohr for all atoms, and an $RK_{max}$ of 9.0 was used.  Here $RK_{max}$ is the product of the sphere radius and the largest basis wave vector.  Approximately 1000 $k$-points in the full Brillouin zone were used for the self-consistent calculations run to convergence and approximately 10,000 points were used for the transport calculations.  The internal coordinates were relaxed using the standard GGA until forces were less than 2 mRyd/bohr. 
All other calculations, the relaxations excepted, included spin-orbit.

\begin{figure} [h!]
\includegraphics[width=7cm]{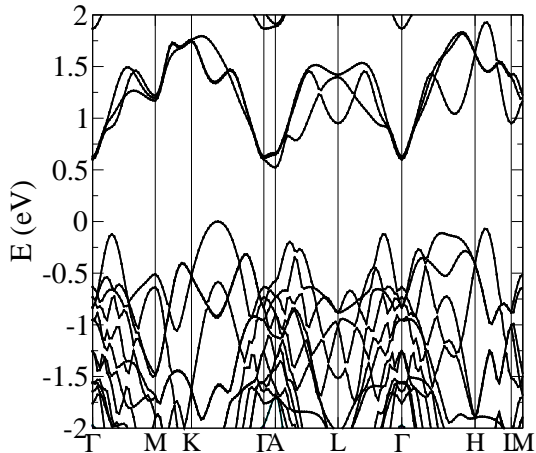}
\caption{  The calculated band structure of hexagonal AgBiSe$_{2}$.}
\end{figure}
Figure 4 depicts the calculated band structure of hexagonal AgBiSe$_2$.
From the figure, we see that this material is a semiconductor of band gap 0.52 eV, with conduction band minimum at the A point and a valence band maximum located between the $\Gamma$
and K points (the Brillouin zone for these points is found in  Ref. \onlinecite{lax}).   The valence band, in addition contains substantial degeneracy, with at least four subsidiary maxima less than 100 meV from the band edge.  These maxima all originate from the 
same general Fermi surface structure, and in addition we note a wide range of masses for these maxima, with the maximum between $\Gamma$ and H having the heaviest mass and the adjacent maximum 
between H and L having the lightest.  This accords with the discussion in the Introduction regarding the anisotropy in band extrema.
To demonstrate the potential of this material for high $ZT$, in Figure 5 we present a plot of the 300 K  calculated power factor vs. carrier concentration $\sigma/\tau$ and compare with $p$-type Bi$_2$Te$_3$, the highest performing room temperature thermoelectric.  Although the $p$-type thermopower and conductivity of AgBiSe$_2$ are nearly isotropic, Bi$_2$Te$_3$ has substantial anisotropy, as is well known.  To ensure a fair comparison we present the conductivity-averaged quantities (see Ref. \onlinecite{chen} for details on these quantities) since in a typical experiment a polycrystalline sample is used, which tends to average the transport over the principal axes.  

The figure depicts $S^2 \sigma/\tau$ results much larger than those of Bi$_2$Te$_3$. In particular, at the doping where the thermopower is 200$\mu$ V/K (indicated by the asterisks), the $S^{2} \sigma/\tau$ of AgBiSe$_2$ is {\it double} that of Bi$_{2}$Te$_{3}$.  This doping, which is p= $2 \times 10^{19}$ cm$^{-3}$ for Bi$_{2}$Te$_{3}$ and $1.2 \times 10^{20} cm^{-3}$ for AgBiSe$_2$, is approximately the doping of optimized ZT for \BiTe and is probably at or near
\begin{figure} [h!]
\includegraphics[width=6.7cm,angle=90]{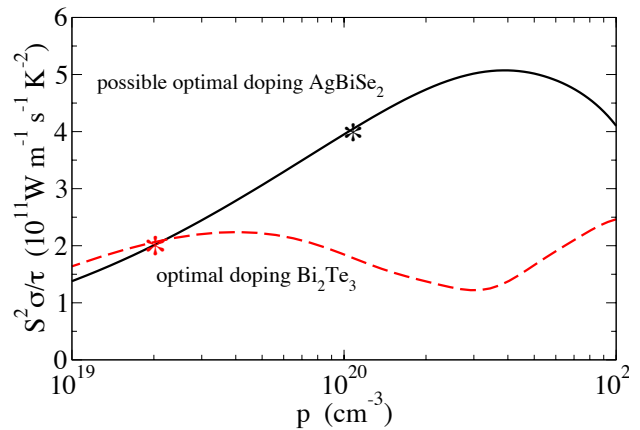}
\caption{  The calculated power factor divided by scattering time, S$^{2}\sigma/\tau$ of $p$-type \BiTe and AgBiSe$_{2}$.  The asterisks indicate the doping where the calculated thermopower is 200 $\mu$ V/K, which is usually near the optimal doping for a high performance thermoelectric.}
\end{figure}
optimal for AgBiSe$_{2}$.  While one cannot exclude scattering time differences from the comparison, this result suggests that AgBiSe$_2$ may show comparable power factors to those of Bi$_{2}$Te$_{3}$, and when combined with the observed low lattice thermal conductivity of AgBiSe$_2$ \cite{morelli}
suggests a high potential for room-temperature thermoelectric performance.

In Figure 6 we depict the calculated thermopower for AgBiSe$_2$ as a function of carrier concentration.  Note that the $p$-type thermopower is virtually isotropic.  We have indicated the range of thermopowers
\begin{figure} [t!]
\includegraphics[width=8cm,angle=90]{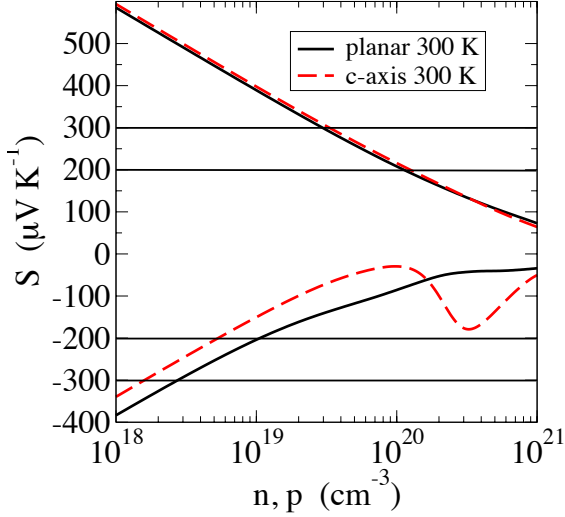}
\caption{  The calculated thermopower of hexagonal AgBiSe$_{2}$ at 300 K.}
\end{figure}
between 200 and 300 $\mu$V/K as this is the general range of thermopower over which $ZT$ is maximized. From this we find optimal 300 K $p$-type doping levels of 3 $ \times 10^{19} - 1.3 \times 10^{20} $cm$^{-3}$.  For $n$-type the optimal doping levels are much lower, indicating a lower likelihood of good thermoelectric performance.

In Figure 7 we present the calculated 300 K electrical conductivity and thermopower ratios $\sigma_c/\sigma_{ab}$ and $S_c/S_ab$.  In view of the great {\it carrier pocket} anisotropy depicted in Figure 2 (bottom), it is of interest that the thermopower and electrical conductivity of $p$-type AgBiSe$_{2}$ vary by no more than 10 percent from the c-axis to the plane.  Note that the electrical \begin{figure} [t!]
\includegraphics[width=8cm,angle=90]{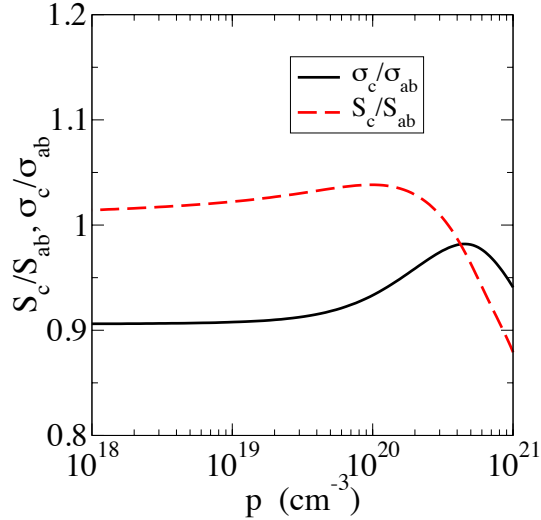}
\caption{  The calculated electrical conductivity  ratio $\sigma_c/\sigma_{ab}$ and thermopower ratio $S_c/S_{ab}$  of hexagonal AgBiSe$_{2}$ at 300 K.}
\end{figure}
conductivity, with respect to an unknown scattering time $\tau$, is calculated as the transport function $N(E)v^{2}_{i}(E)$ integrated with the derivative of the Fermi function, a function sharply peaked around E$_F.$  Here $N(E)$ is the density of states and $v_{i}$ the Fermi velocity in direction $i$.  Hence the anisotropy of the conductivity is a good measure of the anisotropy of the transport function itself.

We note that, as suggested by Figures 6 and 7, in general the overall {\it global} electronic transport anisotropy need not be as large as the {\it local} carrier pocket anisotropy.  Only in the case of a single band extremum are these two anisotropies equal.  When multiple extrema, as depicted in the bottom panel of Figure 2, are present the overall anisotropy can be much less than the carrier pocket anisotropy, due to the varying relative orientations of the band extrema dictated by the crystal symmetry.  Thus despite the great variation in directional effective mass in the hole pockets in the bottom of Figure 2, the overall electronic transport in $p$-type AgBiSe$_{2}$ is very nearly isotropic.

Actual values of the electrical conductivity depend on the electronic scattering time $\tau$, which can vary substantially both from one material to another and by doping level within a particular material.  As is well known, for temperatures above the Debye temperature \cite{allen} $\tau$ is generally inversely proportional to the dimensionless electron-phonon coupling constant $\lambda$, assuming the absence of extrinsic factors such as grain boundary scattering.  While $\lambda$ is not directly available from the first principles calculations we have performed, we estimated above its value as 0.49 from experimental measurements described in Section II.  This section also presented measurements of the thermal conductivity, which allows estimation of the potential $ZT$ of $p$-type AgBiSe$_{2}$.  

As described in Appendix A, from these data we estimate
the 300 K $ZT$ value of optimally doped $p$-type AgBiSe$_2$ as 0.4 - 0.7.   These values consider only optimization of the electronic transport; optimization of the lattice transport can be expected to yield additional performance benefits.  These promising performance values are representative of the positive effects of carrier pocket electronic anisotropy in producing favorable thermoelectric behavior. 

\section{Conclusion}

Carrier pocket electronic anisotropy is seen to positively impact thermoelectric performance, provided that the crystal obeys a minimum symmetry and the relevant band edge is located away from the $\Gamma$ point so that the material can experience the beneficial effects of both band degeneracy and a large effective mass {\it ratio} in the ellipsoidal effective mass tensor.  $p$-type AgBiSe$_{2}$ appears to be a material in which these benefits are present, with ``plate-shaped" Fermi surfaces, a large predicted power factor and ultimately a predicted 300 K $ZT$ value of 0.4 - 0.7, considering only optimization of the electronic properties.  The favorability of this material suggests that other materials with such anisotropic features may exist; indeed, two of the best performing thermoelectrics Bi$_2$Te$_3$ and PbTe, exhibit a great deal of such anisotropy.  Searches for other potentially high performance thermoelectrics with these anisotropic behaviors may therefore be of interest.
\\
\\
\\
{\bf Appendix A : Method for estimation of thermoelectric performance of $p$-type AgBiSe$_{2}$}
\\

Thermoelectric performance is measured as 
\bea
ZT = \frac{S^2 \sigma T}{\kappa}
\eea
Here S is the Seebeck coefficient, $\kappa$ the thermal conductivity, and $\sigma$ the electrical conductivity.  To make quantitative estimation of thermoelectric performance we need the values of these three quantities.  We do this using a combination of our first principles calculations, the experimental data just described, and a few assumptions.

Although in $p$-type AgBiSe$_2$ the Seebeck coefficient and electrical conductivity are calculated to be essentially isotropic, we describe here briefly the application of Eq. (17) in the anisotropic case.  For a single crystal sample, the above expression for $ZT$ directly applies for the Seebeck coefficient, electrical and thermal conductivity measured in a particular direction, leading to a value of $ZT$ which is directionally dependent.  The more typical situation, as in our experimental work, is that of a polycrystalline sample with random grain orientation.  In that case $ZT$ of the sample is isotropic, and the values of S and $\sigma$ which enter the $ZT$ expression are as follows (we assume diagonal $S$ and $\sigma$ tensors):
\bea
S&=& \frac{S_{xx}\sigma_{xx}+S_{yy}\sigma_{yy} + S_{zz}\sigma_{zz}}{\sigma_{avg}}\\
\sigma_{avg} &\equiv& \sigma = \frac{\sigma_{xx}+\sigma_{yy}+\sigma_{zz}}{3}
\eea
Note that the above expression neglects extrinsic effects such as grain boundary scattering that may reduce $\sigma_{avg}$ from the average of the corresponding single-crystal direction-dependent conductivities.  A detailed consideration of the effect of transport anisotropy on $ZT$ may be found in Ref. \onlinecite{bies}.

Returning to calculating the $ZT$ of $p$-type AgBiSe$_2$, our first assumption is regarding optimal doping. Rather than estimate the doping dependence of thermoelectric performance, we note that in most high performance thermoelectrics, the Seebeck coefficient magnitude at optimal doping is between 200 and 300 $\mu$V/K.  The reasons for this are two fold. Firstly, Seebeck coefficients below this range do not permit high $ZT$.  Note that the Wiedemann-Franz relation implies that a minimum thermopower of 156 $\mu$V/K is necessary for a $ZT$ of unity, and this is for a nil lattice thermal conductivity, which is clearly unrealistic.  For example, the thermopower of optimally doped Bi$_{2}$Te$_{3}$ is approximately 200 $\mu$V/K \cite{goldsmid_book}. Secondly, for thermopower values above 300 $\mu$ V/K, the chemical potential is typically in the band gap, and electrical conductivity is correspondingly reduced due to the low carrier concentration.  Since the electrical conductivity we measured in this material is already somewhat low for a high performance thermoelectric (about 1/6th the value for optimized Bi$_{2}$Te$_{3}$ \cite{goldsmid_book}), we will take the higher carrier concentration thermopower of 200 $\mu$ V/ K as representing a sample of likely optimal doping.

Our second assumption concerns the thermal conductivity. This may generally be written as a sum of lattice and electronic thermal conductivity.  The electronic thermal conductivity $\kappa_{electronic}$ can be readily estimated from the electrical conductivity $\sigma$ and the Wiedemann-Franz relation, in which $\kappa_{electronic} = L_0\sigma T$, with L$_0$ the Lorenz number = 2.45 $\times 10^{-8}$ (V/K)$^2$.  This relation is usually accurate for metals and heavily doped semiconductors as considered here.  The lattice thermal conductivity $\kappa_{lattice}$ typically follows a 1/T relation, leading to an estimation of $\kappa_{lattice}$ for AgBiSe$_{2}$ at 300 K of 0.7 W/m-K.

We also make a final assumption regarding the electrical conductivity, in particular the scattering time.  The first principles calculations yield for the $n$-type sample experimentally measured a $\sigma/\tau$ at 300 K of $2 \times 10^{18}$ ($\Omega$ - m - s)$^{-1}$, and given the measured conductivity of 160 S/cm, yield an average scattering time $\tau$ of $8 \times 10^{-15} $s.  To translate this time to a scattering time for optimally doped $p$-type requires consideration of two specific issues: the change in carrier type from $n$-type to $p$-type, and the change in carrier concentration from the $n=2.3 \times 10^{19}$cm$^{-3}$ in our experimental work to the likely optimal $p$-type doping $p= 1.2 \times 10^{20}$ cm$^{-3}$.

Regarding the carrier change from $n$-type to $p$-type, one is really asking about the associated difference in electron-phonon coupling, since such coupling is the basic mechanism of resistivity in heavily doped semiconductors above the Debye temperature.  To help assess this issue,  in Figure 8 we present the calculated density-of-states of AgBiSe$_{2}$. In both the valence and conduction band the Se atoms contribute substantially to the DOS, with the Ag comparatively less in both.  The Bi DOS is more substantial in the conduction band than in the valence band.  Overall, however, the relative proportions of the partial DOS in the conduction and valence bands are similar - both are substantially hybridized  This suggests that the interatomic interactions which produce the electron-phonon coupling may not be too dissimilar from the conduction and valence bands, and 
that, as a first approximation, one may take the electron phonon coupling and associated scattering times to be the same for both bands.
\begin{figure} [t!]
\includegraphics[width=6.5cm,angle=90]{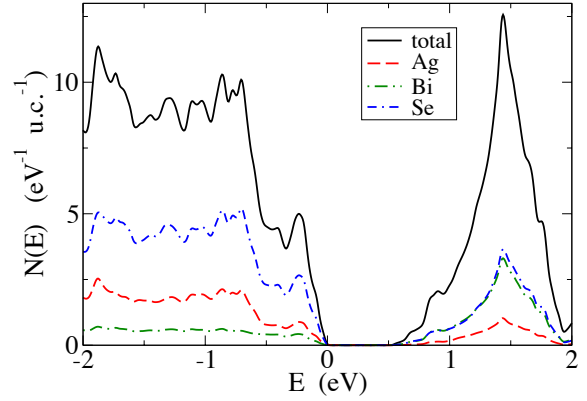}
\caption{  The calculated density-of-states of AgBiSe$_2$.}
\end{figure}
Hence we may estimate the conductivity and $ZT$ of optimally doped $p$-type AgBiSe$_{2}$ from this scattering time.  We find a 300 K conductivity of 800 ($\Omega$-cm)$^{-1}$, approximately 80 percent of the value of optimally doped Bi$_{2}$Te$_{3}$, which yields a 300 K $ZT$ value of 0.7.

At a further level of refinement, we consider that the likely optimal doping level of $p$-type AgBiSe$_{2}$ is substantially higher - more than a factor of five larger - than the experimental carrier concentration in our $n$-type sample, which may affect carrier mobility.  To assess this effect, we examine previous work \cite{parker_PbSe} which finds that in the heavily-doped degenerate limit the carrier mobility $\mu$ typically follows the approximate relationship $\mu \sim p^{-0.6}$.  Taking this same proportionality for the scattering time yields a 300 K conductivity of 300 ($\Omega$-cm)$^{-1}$ and an estimated ZT value of 0.4.

We note that experimental values of $ZT$ will depend on actual scattering times, so that the above should be taken only as first estimates.   However, they demonstrate the potential of $p$-type AgBiSe$_{2}$ as a room temperature thermoelectric.  In fact, the actual performance of this material may well be better than these estimates, because no effort has been made to model reductions of lattice thermal conductivity as are often attained by alloying or nanostructuring.  For example, we may estimate the lattice thermal conductivity
of optimally doped Bi$_2$Te$_3$ from the Wiedemann-Franz relation, its $ZT$ of unity, and Seebeck coefficient and conductivity of 200 $\mu V/K$ and \cite{goldsmid_book}  1000 ($\Omega$-cm)$^{-1}$ as just 0.45 W/m-K, which is likely near the minimum possible for this system, and much lower than the 1.7 W/m-K bulk value \cite{rowe_book}.  We anticipate similar reductions may be available for AgBiSe$_2$; indeed, recent work on AgBiSe$_2$ nanoplates \cite{xiao1} finds lattice thermal conductivity values of 0.45 W/m-K in the pure material and values as low as half this when Sb alloying is considered.
\\
\\
{\bf Appendix B: Demonstration of isotropy of thermopower within effective mass approximation}
\\

We here demonstrate directly that an anisotropic effective mass tensor leads to an isotropic Seebeck coefficient, within a single band model
constant scattering time approximation (CSTA).  Earlier references to this Seebeck isotropy may be found in Refs. \onlinecite{chandrasekhar,kolod}.  We begin with the anisotropic dispersion relation for a single band extremum within the effective mass approximation (we take $\hbar=1$ throughout):
$E_{\bf k} = k_x^2/2m_x + k_y^2/2m_y + k_z^2/2m_z$.
Now, equation (4) in Section IV yields a transport function $\sigma_{\alpha\alpha}$:
\bea
\sigma_{\alpha \alpha}&=& \tau \int \frac{v^2_{\alpha,{\bf k}}}{|v_{\bf k}|}dS
\eea
We now make a scale transformation, writing $k_{\alpha}=\sqrt{2m_{\alpha}}k^\prime_{\alpha}$.
and convert $k^{\prime}_{x},k^{\prime}_{y}$ and $k^{\prime}_{z}$ to spherical coordinates.  Now, the surface area element dS can easily be shown to be
\bea
dS &=& 2E f_1(m_x,m_y,m_z, \theta,\phi) \eea
defining an angular function f$_1$.  Similarly, one finds that
\small
\bea
|v_{\bf k}| &=&  \sqrt{E} g_1(m_x,m_y,m_z,\theta,\phi) \nonumber
\eea
\normalsize
defining a function $g_1$, and
\bea
v^2_{\alpha,{\bf k}} &=& 2 En_{\alpha}^2(\theta,\phi)/m_{\alpha}
\eea
where $n_{\alpha}(\theta,\phi)$ is the direction cosine of $k^{\prime}_{\alpha}$ (i.e., $n=\cos(\theta)$ for $\alpha=z$).
Now we have
\bea
\sigma_{\alpha\alpha}(E) &=& \int v^2_{\alpha}/|v| dS = E^{3/2} \int g_{2,\alpha}(m_x,m_y,m_z,\theta,\phi) d\theta d\phi \nonumber
\eea
where the function $g_{2,\alpha}$ contains all the angular $k^{\prime}$-space dependence and effective mass dependence. The angular integrations
yield a third function $g_{3,\alpha}(m_x,m_y,m_z)$ so that we have
\bea
\sigma_{\alpha\alpha}(E) &=& E^{3/2} g_{3,\alpha}(m_x,m_y,m_z)
\eea
However - {\it and this is the key point} - when inserted into Eq. (1), the effective-mass dependent term $g_{3,\alpha}$ cancels, since
it is present both in the numerator and denominator of Eq. (1).  Hence the Seebeck coefficient is isotropic, even
with an anisotropic effective mass tensor.

Note that although we have assumed the constant scattering time approximation for simplicity, this result holds so long as the scattering time is a function
of energy alone, and not direction.

{\bf Acknowledgment.} This work was supported by the U.S. Department of Energy, Office of Science, Basic Energy Sciences, Materials Sciences and Engineering Division (AFM) and the S3TEC, an Energy Frontier Research Center funded by the U.S. Department of Energy, Office of Science, Basic Energy Sciences (DP and DJS).


\begin{references}
\bibitem{footnote} Note that in this calculation of $m_{\sigma}$ for Bi$_{2}$Te$_3$ we are assuming the use of a polycrystalline sample for which the effective conductivity mass $m_{\sigma}$ is given by $ 3/m_{\sigma} = 1/m_1 + 1/m_2 +1/m_3$. The DOS mass $m_{DOS}$ is unchanged by the use of a polycrystalline sample.
\end{references}
\end{document}